\documentclass[amsmath, twocolumn, prb, superscriptaddress]{revtex4-1}
\usepackage[utf8]{inputenc}
\usepackage{natbib}
\usepackage{graphicx}
\usepackage{subfigure}  
\usepackage{amssymb,amsmath} 
\usepackage{multirow} 
\usepackage{color,soul} 
\usepackage{tabularx} 
\usepackage{lineno} 

\usepackage{changes}

\definechangesauthor[color=red,name={AB}]{AB}
\definechangesauthor[color=green,name={Comments by AB}]{CAB}
\definechangesauthor[color=blue,name={Comments by NM}]{NM}

\begin{document}

\title{Anharmonic Thermodynamics of Vacancies Using a Neural Network Potential}

\author{Anton S. Bochkarev}
\email{anton.bochkarev@cea.fr}
\affiliation{LITEN, CEA-Grenoble, 17 rue des Martyrs, 38054 Grenoble Cedex 9, France}
\author{Ambroise van Roekeghem}
\affiliation{LITEN, CEA-Grenoble, 17 rue des Martyrs, 38054 Grenoble Cedex 9, France}
\author{Stefano Mossa}
\affiliation{Univ. Grenoble Alpes, CEA, CNRS, IRIG, SyMMES, F-38000 Grenoble}
\affiliation{Institut Laue-Langevin, BP 156, F-38042 Grenoble Cedex 9, France}
\author{Natalio Mingo}
\email{natalio.mingo@cea.fr}
\affiliation{LITEN, CEA-Grenoble, 17 rue des Martyrs, 38054 Grenoble Cedex 9, France}
\date{\today}
%
\begin{abstract}

Lattice anharmonicity is thought to strongly affect vacancy concentrations in metals at high temperatures.
It is however non-trivial to account for this effect
directly using density functional theory (DFT).
Here we develop a deep neural network potential for aluminum that overcomes the limitations inherent to DFT, and we use it to obtain accurate anharmonic vacancy formation free energies as a function of temperature. While confirming the important role of anharmonicity at high temperatures, the calculation unveils a markedly nonlinear behavior of the vacancy formation entropy and shows that the vacancy formation free energy only violates Arrhenius law at temperatures above 600 K, in contrast with previous DFT calculations.

\end{abstract}
\maketitle
%
\section{\label{sec:introduction}Introduction}

\textit{Ab initio} calculations based on density functional theory (DFT) have found numerous applications
in materials science. However, due to computational costs, DFT calculations have strong 
limitations when it comes to predicting properties of large systems, studying extended time scales or
computing properties of materials at elevated temperatures. For example, calculating finite temperature phonon densities of states, or vacancy formation energy in solids require performing long Molecular Dynamics (MD) simulations in large supercells.
Machine learning (ML) technniques offer a solution to this problem. Different ML approaches have been used in the attempt 
to overcome the DFT limitations while keeping its accuracy for various applications~\cite{behler_perspective:_2016,behler_generalized_2007,schutt_schnet:_2017,cubuk_representations_2017,bartok_gaussian_2015,kobayashi_neural_2017,rostami_optimized_2018,nyshadham_general_2018,artrith_implementation_2016,takahashi_linearized_2018,Botu2017,shapeev_moment_2016,Natarajan2018,Ward2016,Zong2018,kuritz_size_2018}. 
Over the years, some features of these ML potentials have emerged as necessary for a proper application in materials science. A most relevant one is that the system descriptors should be invariant with respect to rotations and translations of the system as a whole, and with respect to permutations in the ordering of the atom inputs. Another important feature is to be able to train the potential simultaneously with energies and forces, since both are obtained by DFT for the same effort~\cite{pukrittayakamee_simultaneous_2009}. Early NN potentials were only trained to energies, thus discarding the precious information provided by the forces. Other works have trained only to forces, which allows to render properties like the vibrational density of states but is not enough to get formation energies for example~\cite{Kruglov2017,glielmo_accurate_2017,Botu2017}.

In this work, we demonstrate the performance of our ML interatomic potential based on a deep neural network (NN) that is trained to reproduce DFT energies and forces.
To show that the NN potential can successfully overcome the limitations of DFT, we compute the vacancy formation free
energy as a function of temperature in Al. 
Usually the experimental data on vacancy concentrations in metals is analyzed in terms of the Arrhenius law, implying that 
the vacancy formation enthalpy $H_f$ and entropy $S_f$ are assumed to be temperature independent. Deviation from this
law at high temperature in some cases (in the case of Al in particular) was initially explained by the presence of divacancies for which the corresponding formation enthalpy and entropy are also constant with temperature~\cite{Khellaf2002,Seeger1968,Seeger1973,SIEGEL1978117,KRAFTMAKHER199879}. 
Previous theoretical works~\cite{Gilder1975}, simulations based on classical interatomic potentials~\cite{Foiles1994}, and DFT molecular dynamics calculations~\cite{Carling2000}, showed that local anharmonicity significantly affects the monovacancy properties at high temperatures and that this effect alone can explain non-Arrhenius behaviour by the temperature variation of $S_f$ and $H_f$. In the recent work by Glensk et al.~\cite{Glensk2014} the 
vacancy formation free energy was directly computed using \textit{ab initio} thermodynamic integration. The results of this work not only further support the prevalence of monovacancy anharmonicity over divacancies but also suggest that due to anharmonic effects the Arrhenius law is violated at all temperatures. However, 
certain assumptions were made in order to make this calculation possible within a realistic timeframe~\cite{Grabowski2009} (limited number of low precision and small size and time scale calculations were made which then were tuned with the use of high precision and bigger scale reference calculations). 
The NN potential helps to overcome these difficulties as it keeps DFT accuracy but it is much more efficient.
The resulting vacancy formation free energy as a function of temperature obtained in this work is in good quantitative agreement with both the earlier DFT calculations~\cite{Glensk2014} and the available experimental data for the vacancy concentrations~\cite{HEHENKAMP1994907}. However, the character of its temperature variation differs from the earlier DFT result, leading to different conclusions.

\section{\label{sec:Meth}Methods}
\subsubsection{Neural network potential}
The task of the NN potential is to find the relation between the atomic positions and chemical composition of a certain crystal structure, 
and the corresponding energy and forces.
Here, this is done by representing each atom in the structure by an n-dimensional descriptor vector $\textbf{D}$ which includes information 
about the neighbors of the atom within a certain cutoff distance $R_{cut}$. 
For a given atom $i$ and its neighbors $j$ the components of the vector $\textbf{D}_i$ contain information about the neighbor density
around certain regions of space and are computed as:

\begin{equation}\label{eq:Descr}
D^l_i = \sum_j c^l(Z_i)c^l(Z_j) \exp(-\sigma^l(r_{ij}-\eta^l)^2) f_{cut}(r_{ij}, R_{cut}).
\end{equation}

Here, $c^l(Z)$ is the component of the vector representing the chemical identity of an atom defined by its atomic number $Z$. Identifying the atomic type with a vector is done in analogy
to word embedding in natural language processing~\cite{word2vec}. These vectors are initialized randomly and optimized during the training
process. 
We use a Gaussian expansion of the interatomic distances $r_{ij}$ to probe the various regions of space.
The parameter $\eta$ is responsible for choosing the probing region and $\sigma$ defines the `resolution' of the probe. While such an expansion has been a widely used approach since its introduction,~\cite{behler_generalized_2007} the optimization of the parameters $\eta$ and $\sigma$ is rarely performed.
Here, both of these parameters are optimized during the training process.
This allows the NN to automatically select the regions of space that are more important 
for distinguishing atomic configurations, thus freeing from the need 
to manually select the features. 
The third term in Eq.~\ref{eq:Descr} is the cutoff function which ensures that the contribution from the neighbours smoothly goes to zero 
as the distance $r_{ij}$ approaches the cutoff distance $R_{cut}$. Here we use the cosine function:

\begin{equation}\label{eq:Cut}
f_{cut}(r_{ij}, R_{cut}) = \frac{1}{2}(1+cos(\frac{\pi r_{ij}}{R_{cut}})).
\end{equation}

The resulting n-dimensional vector representing each atom is by construction invariant 
upon
translations and rotations of the crystal,
as it depends on the relative interatomic distances. 

Each of those vectors is passed to the deep fully connected neural network such that the output is said to be the atomic energy %
for atom $i$,
$E_i$.
The total energy $E$ of the crystal is obtained by the summation of the atomic energies:

\begin{equation}\label{eq:Etot}
E = \sum_i E_i.
\end{equation}

The use of deep NN in combination with parameterized descriptors construction allows the NN to learn the best representation
automatically, thus reducing the number of hyperparameters that are controlled manually.

To obtain the forces we take the derivative of the NN predicting the energy with respect to the atomic positions:

\begin{equation}\label{eq:Force}
\textbf{F}_{i} = -\frac{dE}{d\textbf{x}_i}.
\end{equation}

The training of the NN is performed by minimizing the combined loss function which simultaneously includes squared losses of energy and forces:

\begin{equation}\label{eq:loss}
\mu||E^{\prime} - E||^2 + \sum_i^{\#\, atoms}||\textbf{F}^{\prime}_i-\textbf{F}_i||^2 \rightarrow min,
\end{equation}

where $E^{\prime}$ and $E$ ($\textbf{F}^{\prime}_i$ and $\textbf{F}_i$) are the target and the predicted energy (forces) respectively.
The parameter $\mu$ is used to give more importance to one of the terms, thus allowing to tune the prediction performance.
We perform the training with the mini-batch stochastic gradient descent using the ADAM optimizer.~\cite{ADAM, Ruder2016}
The total list of optimized parameters include the weights of the neural network, as well as parameters $c^l$, $\sigma^l$ and $\eta^l$ in Eq.~\ref{eq:Descr}.
Simultaneous training with both the energies and forces is beneficial as it allows for a more efficient utilization of the training data.

\subsubsection{Potential training and testing}

We perform \textit{ab initio} molecular dynamics simulations with two different functionals to produce training data for our potential. Computational details are given in the next section.
The training data consists of calculations of $3\times3\times3$ Al conventional fcc supercells with different lattice parameters.

We use 9 different lattice parameters in the range of 3.9 to 4.2 $\AA$ for PBEsol and in the range of 3.95 to 4.25 $\AA$ for PBE. For each of them there are two types of supercells: with and without vacancy. The perfect and defected supercells contain 108 and 107 atoms respectively.

The MD simulations are performed in the NVT ensemble at 3 different temperatures: 100 K, 500 K and 900 K.
In total, each data set consists of 11000 configurations which are randomly split into the training set (9500 configurations) and test set (1500 configurations).
\begin{figure}[b!]
  \begin{center}
    \includegraphics[width=80mm]{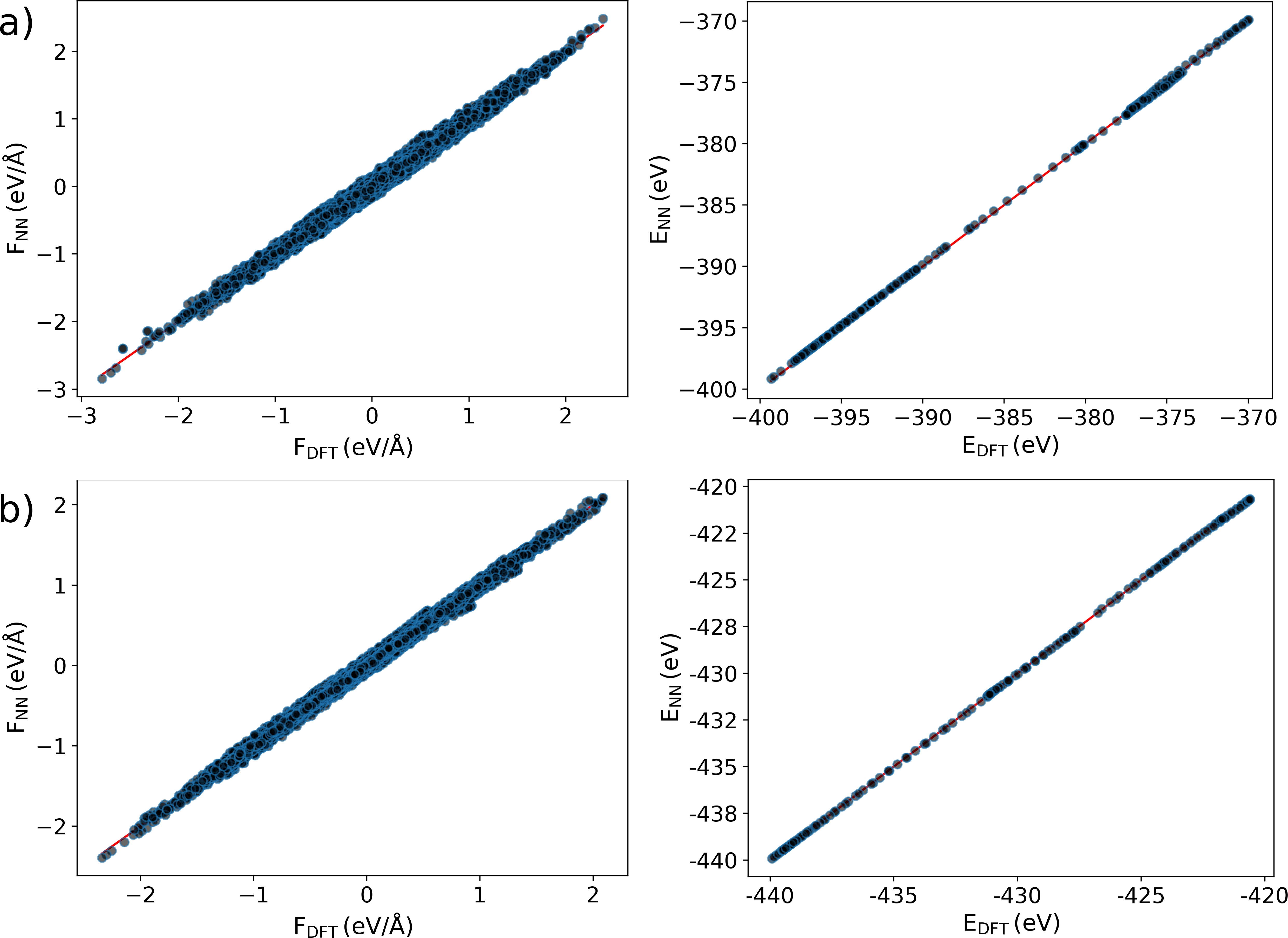}%
  \end{center}
  \caption{Comparison of the predicted energies and forces for the validation data set for a) PBE and b) PBEsol functional.}
  \label{fig:valid}
\end{figure}
Separate trajectories were computed to create a validation data set. In these calculations we use lattice parameters and temperatures which are not presented in the training or test data sets. But for the PBEsol functional the lattice parameters for the validation set are taken within the range of the training set while in the case of PBE the values are taken outside this range. This is done in order to test the ability of the NN to generalize beyond the range of training data. In total, we used 160 configurations in the validation set for PBEsol and 270 configurations for PBE.

In the configuration of the NN that we use, the initial descriptor vector $\textbf{D}$ has dimensionality $n=100$, which is then followed by 6 hidden layers. The first 5 hidden layers have 1000 nodes each and the last one has 500 nodes. The cutoff distance $R_{cut}$ for the cutoff function $f_{cut}$ is set to 8 $\AA$. Table \ref{table:errors} shows the results of the training of the NN potential for different data sets. One can see a very good accuracy in predicting energies and forces for both test and validation sets. 
\begin{table}[tbp!]
\caption{Root-mean-square error (RMSE) and mean absolute error (MAE) of the NN potential for predicting DFT energies and forces.}
 \begin{tabularx}{\columnwidth}{llXX|XX}
\hline
\hline 
\multirow{2}{*}{}       & \multirow{2}{*}{Data set} & \multicolumn{2}{l|}{E, meV/atom}  & \multicolumn{2}{l}{F, eV/$\AA$} \\
                        &                           & RMSE           & MAE           & RMSE           & MAE            \\
\hline
\multirow{2}{*}{PBEsol} & Test                      & 0.40           & 0.27          & 0.021          & 0.014          \\
                        & Validation                & 0.35           & 0.25          & 0.025          & 0.017          \\
\hline
\multirow{2}{*}{PBE}    & Test                      & 0.53           & 0.32          & 0.028          & 0.018          \\
                        & Validation                & 1.4            & 1.3           & 0.035          & 0.023   \\
\hline
\hline
\label{table:errors}
\end{tabularx}
\end{table}
As expected, in the case of PBE data the difference between test and validation errors is higher than the one for PBEsol as the PBE validation data is taken outside the training range. However, the accuracy is still very good (of the order of 1 meV/atom for the energies) showing the ability of the model to generalize on the unseen data. The predictions for the validation sets are also shown in Fig.~\ref{fig:valid}. The force errors of our NN potential are below those reported for other Al machine learning force fields~\cite{Kruglov2017, Botu2017}.
We note that the same level of accuracy can be reached using a much smaller amount of training data. In this work, however, we do not aim at studying the optimal way of selecting training data, but rather at using the maximum amount of data at our disposal. 

\begin{figure}[t!]
  \begin{center}
    \includegraphics[width=80mm]{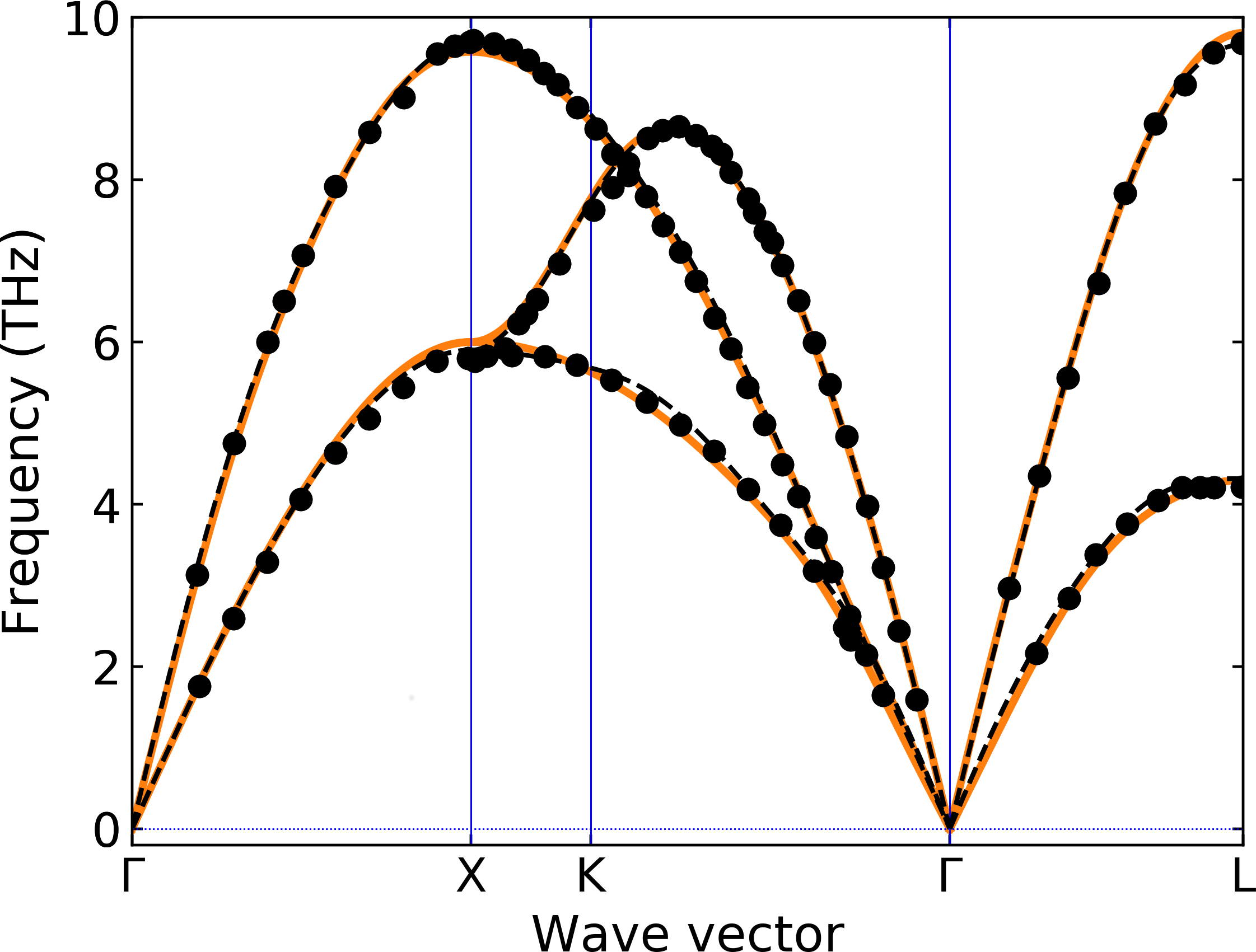}%
  \end{center}
  \caption{Phonon dispersion calculated using the PBEsol NN potential (solid line) compared to the experimental data~\cite{Stedman1966} (symbols) and DFT calculation (dashed line).}
  \label{fig:disp}
\end{figure}

\begin{figure}[pb!]
  \begin{center}
    \includegraphics[width=80mm]{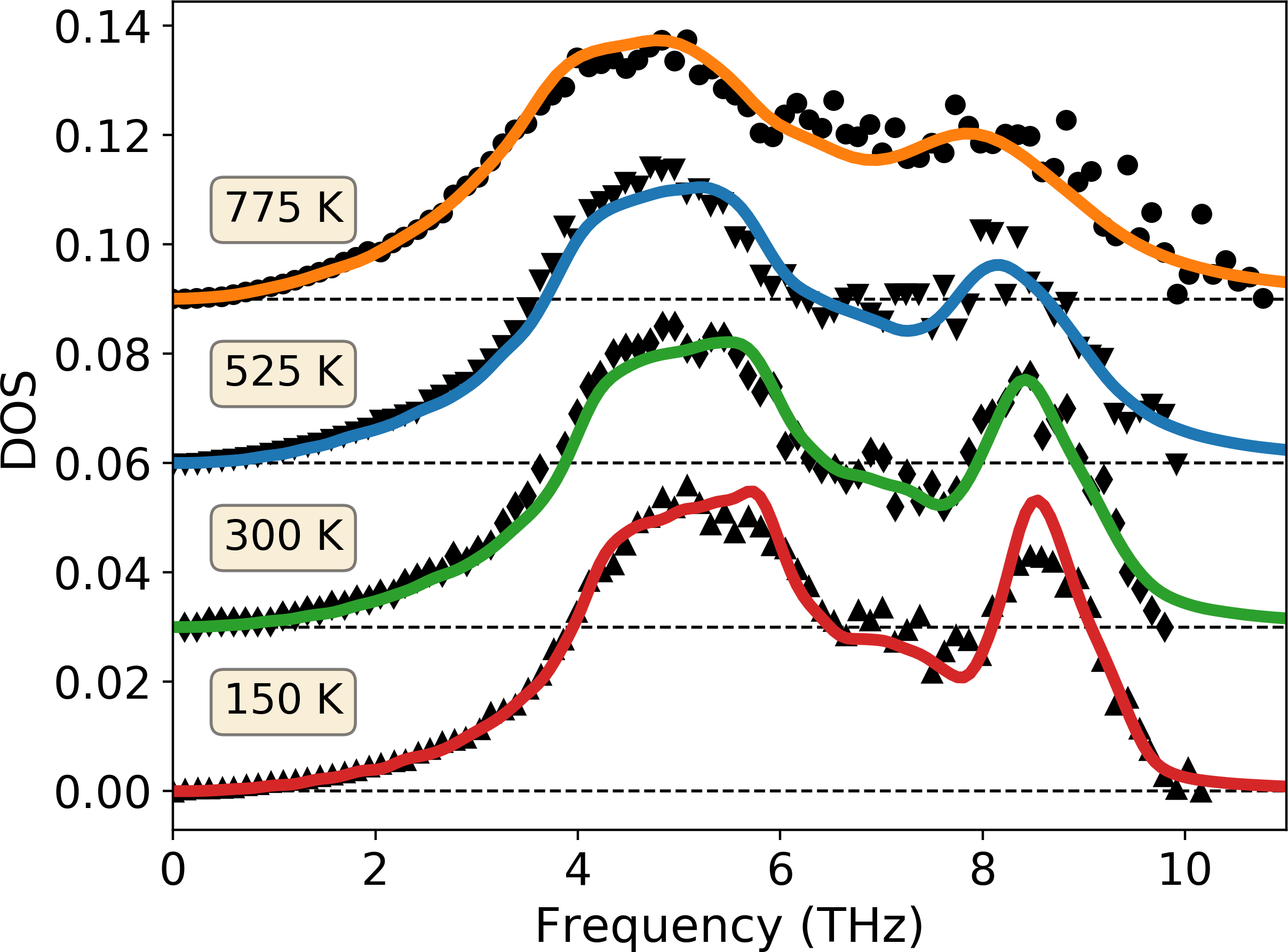}%
  \end{center}
  \caption{Phonon densities of states obtained using the NN potential (solid lines) compared to the experimental data~\cite{Kresch2008} (symbols) at several temperatures. The data for each temperature are shifted along the vertical axis in order to avoid overlapping.}
  \label{fig:DOS}
\end{figure}

To further check the accuracy of the potential we also compute physical properties that can be directly compared to experiment. 
One such property is the low temperature phonon dispersion. Figure~\ref{fig:disp} shows the small-displacements harmonic calculation performed with the PBEsol NN potential, which agrees very well with the experimental data at 80 K~\cite{Stedman1966} as well as with the analogous DFT calculation.
We also compute the anharmonic phonon densities of states at various temperatures through the Fourier transform of the velocity auto correlation function (VACF). To obtain VACF we run NVE MD simulations with 32000 atoms supercells for 4ps at each temperature. The supercells are previously equilibrated in the NVT MD simulation for 4 ps at the corresponding temperature.
Results for the PBEsol NN potential are shown in Fig.~\ref{fig:DOS} together with the experimental data~\cite{Kresch2008}. One can see a very good agreement between calculations and experiment, thus, further confirming the accuracy of the trained potential.
For the PBE NN potential results of these calculations are similar and, therefore, not shown.

\subsubsection{Computational details}
The \textit{ab initio} calculations are performed using 
the Vienna ab initio simulation package.~\cite{Kresse1996, Kresse1996-1}
For exchange and correlation effects, we employ the generalized-gradient approximation using the PBE and PBEsol parametrization.~\cite{PBE,PBEsol}
Ion-electron interactions are described using the projector augmented wave method,\cite{Bloechl1994} with a plane-wave energy cut off of 500 eV.   
A gamma-centered Monkhorst-Pack k-point mesh~\cite{Monkhorst1976} of $8\times8\times8$ is used for Brillouin zone sampling.
The NN architecture and training was implemented using TensorFlow.~\cite{TF} All calculations with the NN potential were performed using ASE.~\cite{asepaper}

\section{\label{sec:results}Results and discussion}
The free energy of a system at given volume $V$ and temperature $T$ can be computed as the sum of two contributions:

\begin{equation}\label{eq:Free}
  G(V,T) = E(V) + G_{vib}(V,T),
\end{equation}

where $E(V)$ is the potential energy of the system, and $G_{vib}(V,T)$ is the vibrational free energy. Calculating this second term is typically 
a very expensive computational task, thus the term is further split into harmonic and anharmonic contributions.
The harmonic part is relatively easy to obtain and it is computed as follows:

\begin{equation}\label{eq:Ghar}
  G_h(V,T) = \int d\varepsilon g(\varepsilon) \left[\frac{\varepsilon}{2}+k_BT\ln\left(1-e^{-\frac{\varepsilon}{k_BT}}\right) \right],
\end{equation}

where $g(\varepsilon)$ is the phonon harmonic density of states obtained for a given volume.
The anharmonic part is usually small compared to the harmonic one but requires significant resources in order to compute it, therefore, it is often ignored. In the case of the
vacancy, however, the anharmonic contributions result in significant changes in formation free energy.
We use thermodynamic integration in order to account for anharmonicity. It allows for the accurate computation of the difference in 
free energy between the reference harmonic and the true anharmonic potential at a given temperature. The free energy difference $\Delta G$ is obtained as follows:

\begin{equation}\label{eq:TI}
  \Delta G(T) = \int\limits_{0}^{1} \left\langle \frac{\partial E(\lambda)}{\partial\lambda} \right\rangle_{\lambda,T}d\lambda.
\end{equation}

Here, $E(\lambda)$ is the potential energy function defined as the sum $(1-\lambda)E_{h} +\lambda E_{ah}$ where $\lambda$ is the coupling parameter
between the harmonic and anharmonic potentials and its values run from 0 to 1. The averaging is performed over the MD run for a given $\lambda$.
The harmonic energy is obtained as:

\begin{equation}\label{eq:Eh}
  E_{h} = E_{0} + \frac{1}{2}\textbf{u}^T\mathbf{\Phi}\textbf{u},
\end{equation}

\begin{figure}[pt!]
  \begin{center}
    \includegraphics[width=80mm]{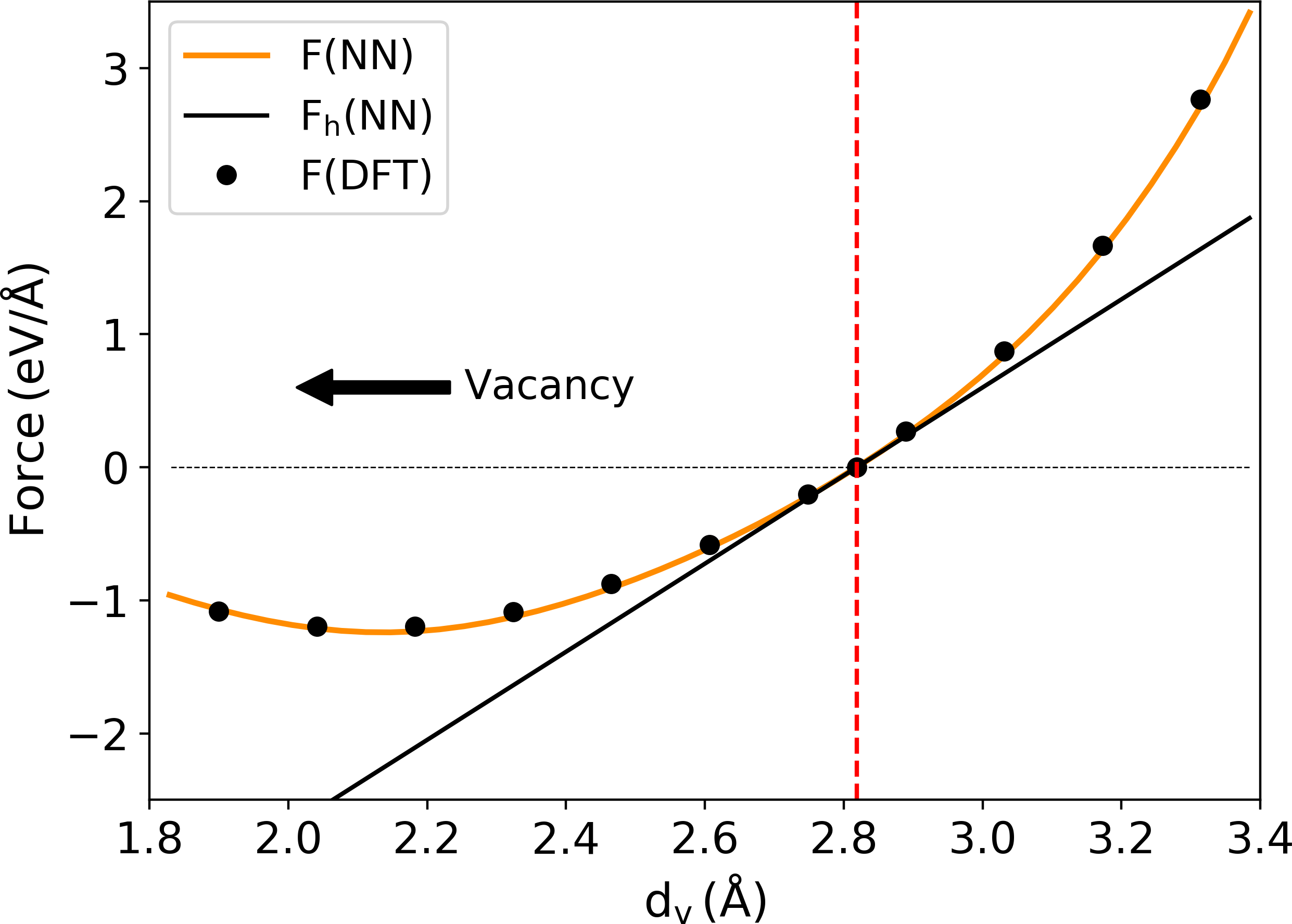}%
  \end{center}
  \caption{Force acting on an atom in the vicinity of vacancy as a function of its distance $\mathrm{d_v}$ from the vacancy. The vacancy is thought to be located at the position of the atom which was taken out to create this vacancy. Symbols show the DFT results and orange line is the NN potential results. Positive sign correspond to the force directed towards the vacancy. Vertical line shows the equilibrium position of the atom. The harmonic force $\mathrm{F_h}$ is obtained by differentiating Eq.~\ref{eq:Eh} and shown with black line for comparison.}
  \label{fig:vac_frs}
\end{figure}

where $E_{0}$ is the energy of the system at equilibrium, $\textbf{u}$ is the vector of atomic displacements from equilibrium and $\mathbf{\Phi}$ is 
the matrix of force constants. The anharmonic part is directly provided by the NN potential.
To illustrate how the NN potential describes the anharmonic behaviour near the vacancy, we gradually displace an atom in the vicinity of vacancy from its equilibrium position in the direction from and towards the vacancy and for each configuration we compute forces acting on the atom with DFT and with the NN potential. It can be seen in Fig.~\ref{fig:vac_frs} that the forces exhibit strong anharmonic behaviour and the NN potential is accurately reproducing the DFT results.

\begin{figure}[pt!]
  \begin{center}
    \includegraphics[width=80mm]{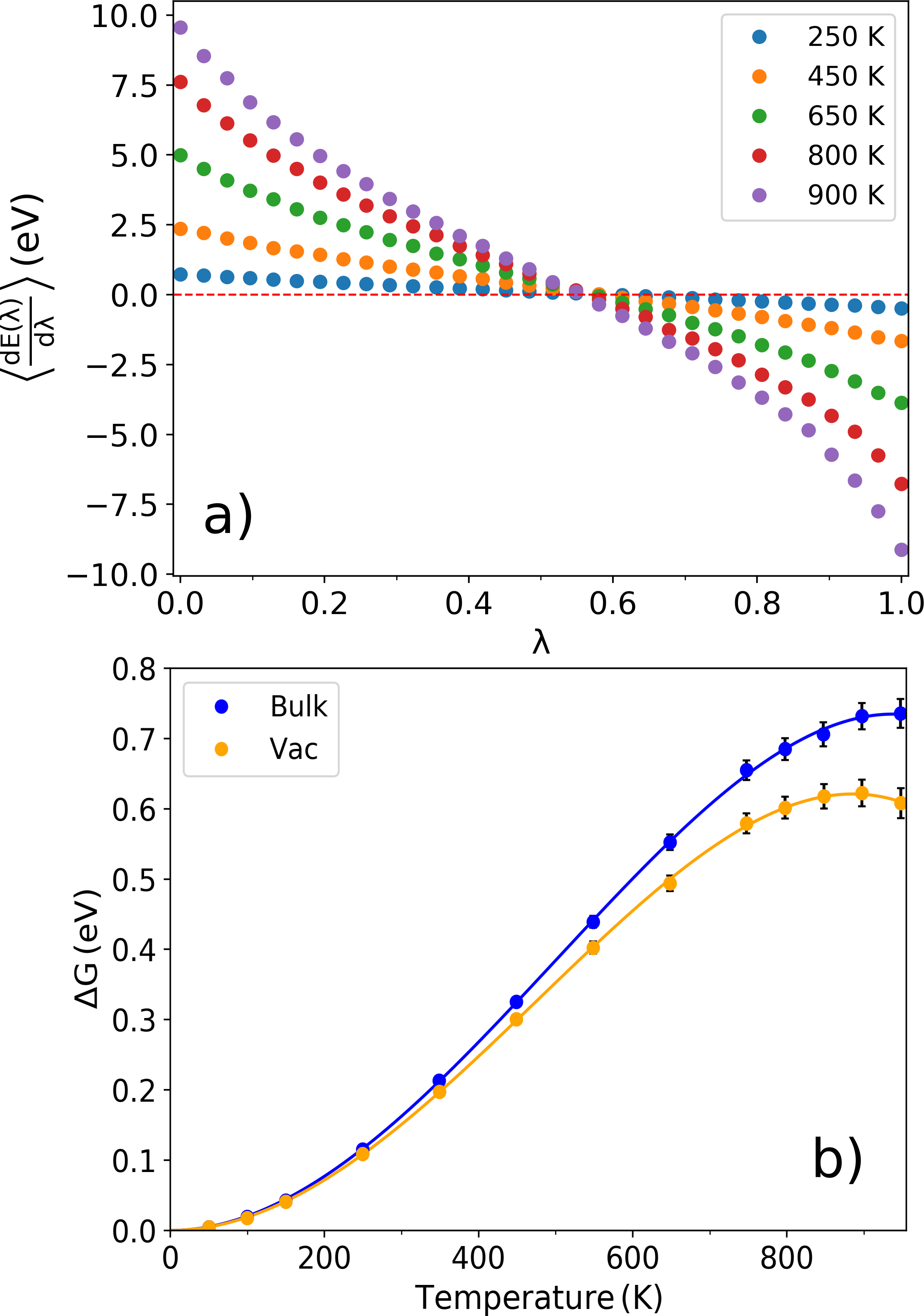}%
  \end{center}
  \caption{Example of the results from thermodynamic integration obtained for the cell with lattice parameter 4.156 $\mathrm{\AA}$. (a) Average of the energy derivative with respect to coupling parameter $\mathrm{\lambda}$ for a given value of $\mathrm{\lambda}$ at selected temperatures for a defected cell. The standard deviation is computed with the coarse-grain sampling.~\cite{Haile1992}
  Corresponding error bars are smaller than the size of the symbols at this scale. (b) Free energy difference between harmonic and anharmonic potential as a function of temperature. Symbols show the integral of the corresponding data on panel (a) and lines are the polynomial fit.}
  \label{fig:ti}
\end{figure}

Knowing the free energy of the supercells with and without vacancy one computes the formation free energy as: 

\begin{equation}\label{eq:Gf}
   G_f(T) = G_v(T)-\frac{N-1}{N}G_b(T),
\end{equation}

where $N$ is the number of atoms in the supercell without vacancy.
The free energy of a supercell at a given temperature is computed in the following way:
(i) First, we compute the harmonic vibrational free energies using Eq.~\ref{eq:Ghar} for a set of 5$\times$5$\times$5 supercells (500 atoms) at 14 different volumes (lattice parameters).
(ii) Next, for each volume the anharmonic corrections $\Delta G(T)$ are obtained from Eq.~\ref{eq:TI} at 13 different temperatures.
At each temperature, integration in Eq.~\ref{eq:TI} is carried out on a mesh of 32 values of $\lambda$ and for each
$\lambda$ the averaging is performed over MD trajectories of about 30 ps with a time step of 1 fs. An example 
of the results of these calculations is shown in Fig.~\ref{fig:ti}.
\begin{figure}[t!]
  \begin{center}
    \includegraphics[width=80mm]{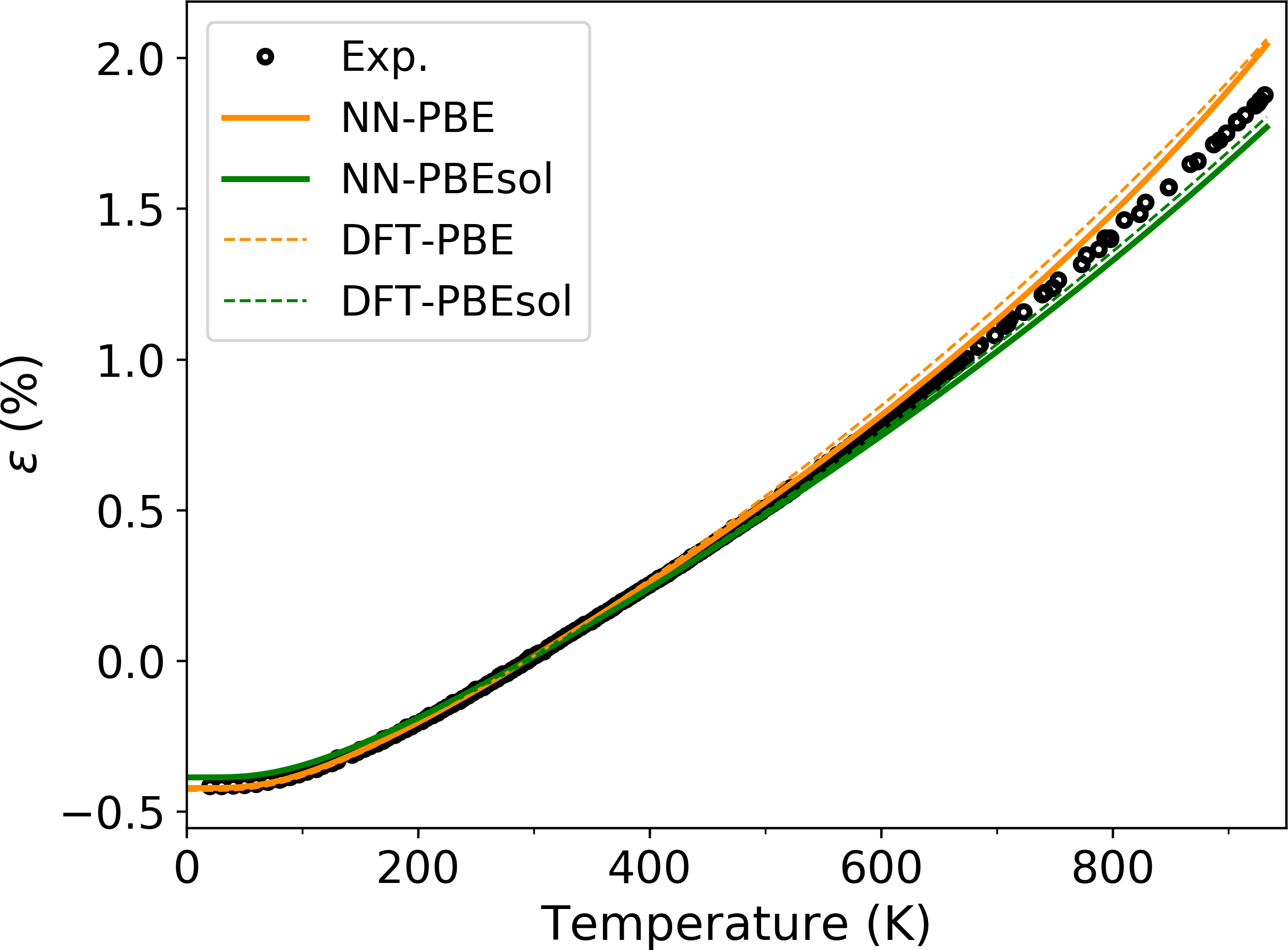}%
  \end{center}
  \caption{The anharmonic thermal expansion of Al computed with the NN potential (solid lines) trained on the data obtained with different DFT functionals compared
          to the experimental data~\cite{touloukian1977} (symbols), and to the quasi-harmonic DFT calculations (dashed lines).}
  \label{fig:Expansion}
\end{figure}
The resulting temperature dependence of $\Delta G(T)$ is fitted to a 4th order polynomial.
(iii) Finally, for a given temperature the free energy $G(V)$ is fitted with the Birch-Murnaghan equation of state~\cite{Birch1947}
to find the lowest energy volume, thus obtaining the temperature dependence of free energy $G(T)$ as well as the volume $V(T)$.
This is done for the systems with and without vacancy, and also with two different NN potentials which are trained using 
independent DFT calculations with different GGA functionals - PBE and PBEsol.
The temperature dependence of volume in terms of the thermal linear expansion is shown in Fig.~\ref{fig:Expansion} together with the experimental and quasi-harmonic DFT data. 
The PBE results overestimate the experimental data at high temperature while PBEsol underestimate. The anharmonic contributions included in the NN potential calculations slightly lower the thermal expansion at high temperature as compared to the DFT quasi-harmonic results. 

\begin{figure}[pb!]
  \begin{center}
    \includegraphics[width=80mm]{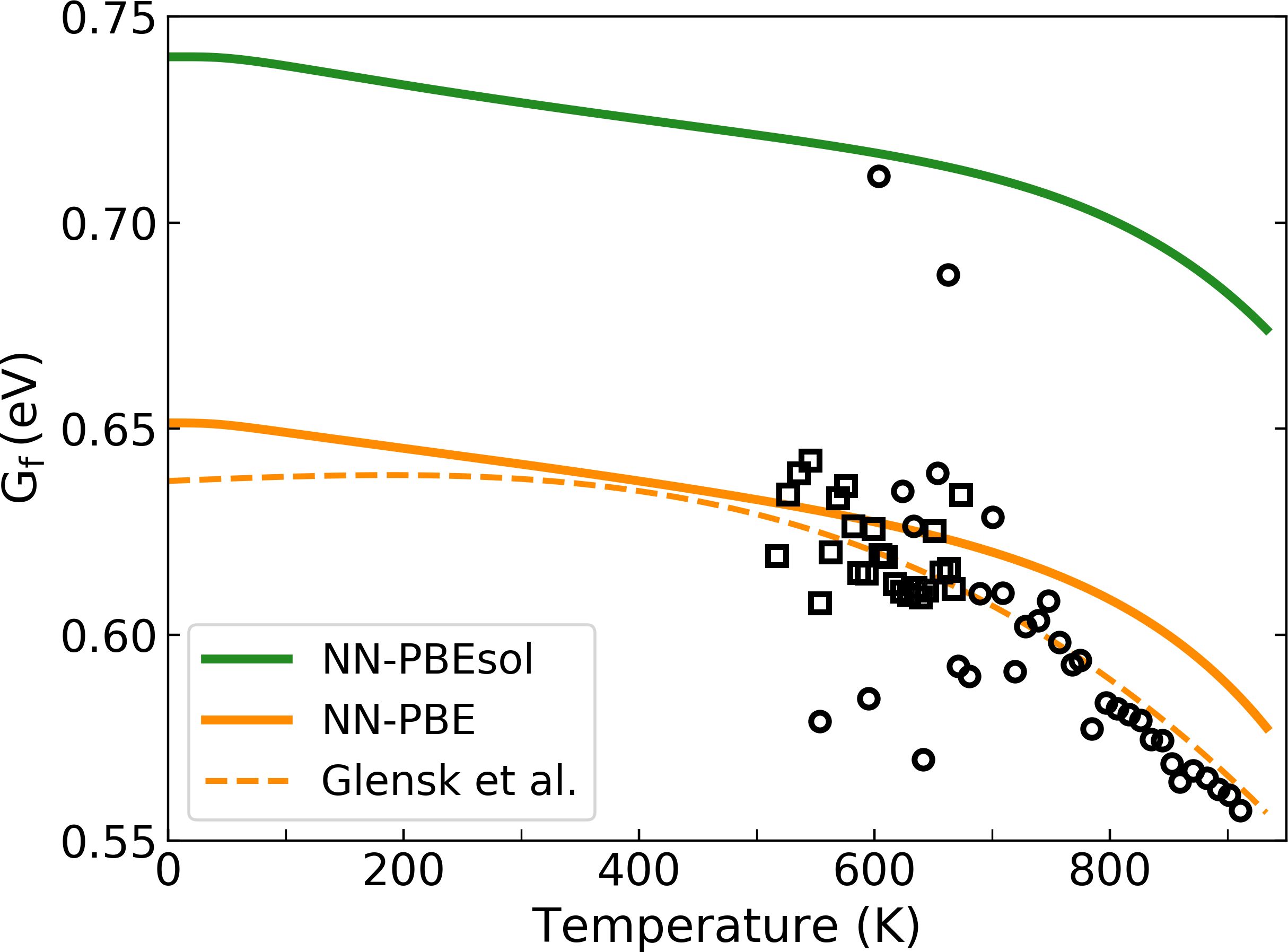}%
  \end{center}
  \caption{Vacancy formation free energy as a function of temperature obtained with the NN potential trained using PBEsol  (green solid line) and PBE data (orange solid line). Dashed line shows the DFT PBE results from reference~[\onlinecite{Glensk2014}]. Symbols represent the experimental data~\cite{HEHENKAMP1994907} for vacancy concentrations obtained from positron lifetime measurements (squares) and differential dilatometry (circles). The formation free energy $G_f$ is calculated from concentrations $c_v$ using relation $G_f = -k_BT\ln c_v$.}
  \label{fig:vacf}
\end{figure}

Figure~\ref{fig:vacf} shows the resulting vacancy formation free energies ($G_f$) as a function of temperature together with
the experimental data~\cite{HEHENKAMP1994907}.
It is important to note that the two sets of results obtained with different functionals, i.e. two different NN potentials trained with independent data sets, exhibit similar variation with temperature but are shifted with 
respect to each other along the energy axis, with the PBEsol result showing larger values.
The PBE result is very close to the experimental data while PBEsol values are too high. This is in agreement with previous calculations of $G_f(T)$ by Glensk et al.~\cite{Glensk2014} for different functionals. The shift is also consistent with the 0 K computations of the vacancy formation energy performed with various DFT functionals~\cite{MEDASANI201596}.

\begin{figure}[t!]
  \begin{center}
    \includegraphics[width=80mm]{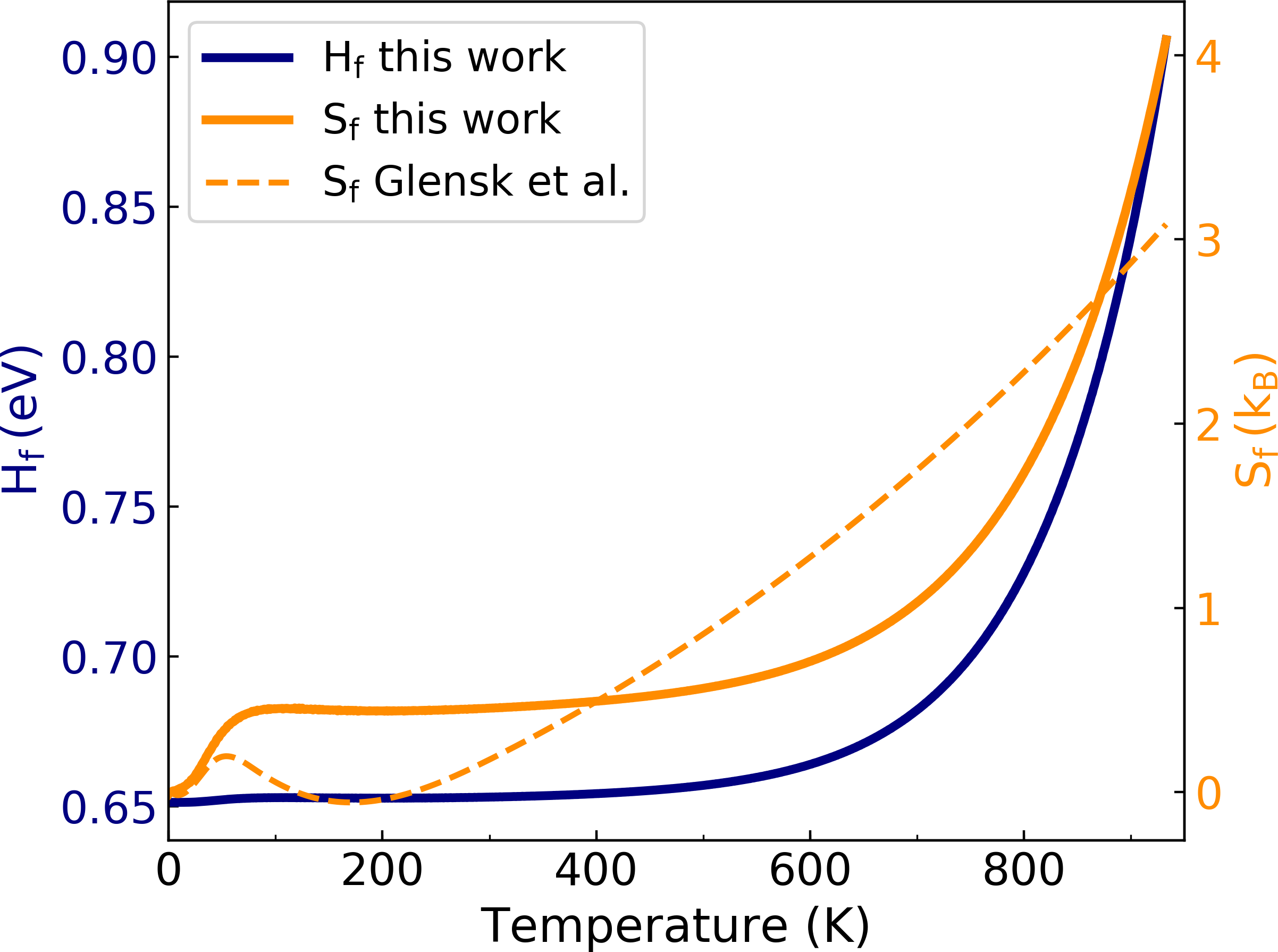}%
  \end{center}
  \caption{Vacancy formation entropy and enthalpy as a function of temperature. Solid lines
  show the results obtained in this work. Entropy is obtained as the derivative of free energy (Fig.~\ref{fig:vacf}) $S_f = -dG_f/dT$ and enthalpy is computed using the relation  $H_f = G_f +TS_f$. The dashed line shows the data from reference~[\onlinecite{GlenskPhD}].}
  \label{fig:vacsf}
\end{figure}

Overall, our PBE results are very close to those of Glensk et al. (also shown in Fig.~\ref{fig:vacf} for comparison), 
i.e. $G_f$ is decreasing with increasing temperature,
and the difference in the absolute values is at most about 0.02 eV, 
mainly due to the offset at 0K. However, a significant difference is found for the vacancy formation entropy $S_f = -dG_f/dT$. Our results for $S_f$ are shown in Fig.~\ref{fig:vacsf} together with the vacancy formation enthalpy $H_f = G_f +TS_f$.
As one can see, the formation entropy and enthalpy are nearly constant in a large temperature range and start to rapidly
increase around 600 K. This is consistent with the widely experimentally used phenomenological Arrhenius law, which assumes $S_f$ and $H_f$ to be temperature independent. Thus, deviation from this assumption at high temperatures can be attributed to anharmonicity.
On the other hand, Glensk and co-workers predict that the
Arrhenius law is not satisfied at 
any temperature.
They find that the entropy in the case of Al first becomes negative and then grows in a linear fashion~\cite{GlenskPhD} (also shown in Fig.~\ref{fig:vacsf} for comparison). Such a behavior is then also attributed to anharmonic effects. 
It is however unlikely that in metals anharmonicity produces such a strong  
effect at low temperatures. The negative sign of the formation entropy around 190K also appears unphysical. The origin of the observed behavior is most likely in the insufficient amount of data points when fitting the anharmonic contributions from Eq.~\ref{eq:TI} as a function of temperature. This means that such a remarkable agreement between the calculated
$G_f$ in reference~\cite{Glensk2014} and the experimental data is somehow accidental.
This, however, does not negate the important conclusion that the anharmonicity 
strongly affects the vacancy thermodynamics at high temperatures, and that
the anharmonicity of monovacancies is enough to explain the experimentally observed trends of vacancy concentrations.

To summarize, in this work we presented a deep neural network interatomic potential. The architecture of the NN is very simple and only based on the interatomic distances. Nevertheless, we demonstrated a high accuracy of the generated interatomic potential. We employed this potential to calculate vacancy formation free energy as a function of temperature. The accuracy and efficiency of the potential allowed us to obtain results which refine the state-of-the-art theoretical description of vacancy thermodynamics in metals. Namely, our results predict that $G_f$ as a function of temperature follows the Arrhenius law at low temperatures, 
in contrast
with previous calculations, and that anharmonicity causes deviation from it only above 600 K.
The results displayed in this paper are for a monatomic system, but our NN potential is also universal in terms of structure and chemical composition.
Applications of this potential for more complex systems will be demonstrated in future works.

\begin{acknowledgments}

This work was supported by Agence Nationale de la Recherche through Carnot projects Si premium and MAPPE, and HPC resources from GENCI-TGCC (project A0030910242).
S.M. is supported by ANR-18-CE30-0019 (HEATFLOW).
\end{acknowledgments}


%

\end{document}